# The Singularity Warfare:
# The metatheoretical Framework

Ridvan Bariiev, PhD

University of Warsaw, Poland

ridvanbari@gmail.com

## Abstract

This paper introduces the "Singularity Warfare" concept, arguing that the accelerating pace of technological revolution, driven by artificial intelligence and quantum mechanics, is fundamentally reshaping the nature of conflict. Moving beyond traditional "Newtonian" warfare and current military doctrines, this framework posits that future battlefields will be defined by a merger of physical and abstract domains, where human imagination and algorithmic logic become a unified, actionable reality. Victory will hinge on a unit's ability to maintain cognitive and technological "coherence" while creating "decoherence" in the adversary. The paper synthesizes theories from physics, philosophy, and futurology to provide a metatheoretical framework for understanding this paradigm shift.

## Introduction

Following the Second World War, modern warfare was traditionally divided into two primary categories: strategic and conventional forces. However, in recent years, a new triad has emerged—cyber, space, and drone forces—marking a significant transformation in the character of warfare. Notably, these three domains have appeared simultaneously within the same historical period, a convergence unprecedented in the history of armed conflict. The integration of these capabilities into a unified operational framework has led to what is now commonly referred to as Multi-Domain Operations (as conceptualized by NATO), Fourth or Fifth Generation Warfare, and Mosaic Warfare (a term used by the United States). More broadly, since the 1990s, these developments have been encompassed under the concept of Network-Centric Warfare. This terminology emerged as an urgent response to the technological changes that began to unfold in the 1980s and 1990s, as advanced innovations were increasingly adapted to military purposes.

These theories primarily offer an engineering perspective on the future of warfare rather than presenting a comprehensive architectural framework. The engineering-centric vision continues to dominate much of the intellectual discourse in the West. For example, Michael E. O'Hanlon, in his 2015 work, illustrates this tendency by emphasizing - "The basic human need to walk the battlefield and to get to know the population in order to conduct proper counterinsurgency operations proved as timeless as ever."[1] He believes that the ongoing technologic changes that impact on the war are going to be significant but certainty can't be called revolutionary.[2] Thus, for O'Hanlon, writing in the 2010s, global military developments did not appear to be moving in the direction of a true Revolution in Military Affairs.

---

[1] Michael E. O'Hanlon, "The Future of Land Warfare, Bloomsbury Publishing PLC, 2015, p. 19
[2] Ibid., 166

A significant portion of scholars continue to resist conceptualizing future warfare through the lens of science fiction, despite its often striking accuracy in anticipating technological and societal developments. One notable example is Robert Heinlein's Starship Troopers, which offers a remarkably prescient depiction of the future soldier. This is not to suggest that we should uncritically adopt every element proposed by the science fiction canon; rather, these works remain among the few intellectual models that offer a meaningful framework for understanding the trajectory of both military and civilian evolution. The next important component of this argument involves examining the intellectual legacy of scholars who sought to construct models capable of addressing the implications of the scientific revolution—an upheaval that is already underway.

However, there is a strong likelihood that none of the existing conceptual models will be sufficient to account for the rapidly accelerating and unpredictable components of future warfare—elements that are emerging with the force and unpredictability of an avalanche. It is reasonable to assert that the transformations currently underway will shape the character of warfare for the next century. Unlike previous historical epochs, the forthcoming Era of Singularity possesses a far more complex nature, with profound implications for both the conduct of war and the conceptual foundations of warfare itself. Accordingly, I will now present my theoretical argument for why we are entering the Era of Singularity—an epoch that is poised to fundamentally alter our understanding of both the theory and the practice of war.

Theoretically, it may be argued that traditional warfare, as it evolved over millennia, was largely grounded in the principles articulated by the Greek philosopher Aristotle. However, in the 21st century—and likely throughout the decades and centuries to come—we are witnessing a gradual but profound transition toward a metaphysical ethos of war. Classical warfare, rooted in tangible domains such as territory, manpower, and kinetic force, aligns with Aristotle's empirical and teleological worldview: war functioned as a means to achieve discernible ends—political power, security, or economic advantage—based on observable causes and hierarchically structured organization. By contrast, contemporary warfare increasingly unfolds within abstract and non-material domains—cyberspace, cognitive space, and algorithmic ecosystems—that resonate more closely with Plato's metaphysical emphasis on forms, ideals, and unseen forces shaping reality. In modern conflict, the primary objective often revolves around dominating perception, controlling narrative, and influencing intent rather than securing physical space alone. Emerging concepts such as cognitive warfare, algorithmic deterrence, and supremacy in the information domain illustrate a paradigmatic shift: war is evolving into a contest over ideas, illusions, and immaterial structures—arguably signaling a transition from an Aristotelian to a more Platonic conception of war.

In other words, the physics of war is evolving from what might be termed Newtonian warfare into an entirely new paradigm. Rooted in classical mechanics, Newtonian warfare represents the traditional framework of conflict—one in which forces, masses, and velocities are predictable and governed by linear cause-and-effect relationships. This model is characterized by clearly defined frontlines, tangible objectives, and an emphasis on physical dominance, force projection, and attrition. It reached its apogee during the Industrial Age, particularly in the large-scale conflicts of the World Wars. Strategically, it is structured around measurable variables and material assets, making it inherently mechanistic in both planning and execution. However, the technologies currently under development for future conflicts are increasingly oriented toward the cognitive domain. In this emerging model, the human mind—or its artificial analogues through AI—will supersede hardware as the dominant force in warfare. The centrality of mind and consciousness will define the core dynamics of the coming Era of Singularity. Crucially, these cognitive capacities need not remain exclusive to human beings.

The most appropriate term to conceptualize the ongoing transformation across technological, societal, and civilizational paradigms is the notion of Technological Singularity, popularized by Ray Kurzweil. However, the origins of the term predate Kurzweil and can be traced back to the classical scientific era that laid the intellectual foundations for the very concept of singularity. It was during the mid-20th century—a period marked by rapid scientific and technological advancement—that the idea began to take shape. The Polish-American physicist Stanislaw Ulam, in an article reflecting on the work of John von Neumann, recounts that von Neumann had remarked "the ever-accelerating progress of technology and changes in the mode of human life, which gives the appearance of approaching some essential singularity in the history of the race, beyond which human affairs, as we know them, could not continue."[3]

In 1983, the seminal manifesto "The Coming Technological Singularity" by Vernor Vinge was published. In this article, Vinge argued that the world was approaching a transformation as profound as the emergence of human life itself—a shift driven by the creation of entities possessing intelligence greater than that of humans. This superior intelligence, he predicted, would exponentially accelerate technological and social progress. Vinge termed the moment of such creation, and the rapid cascade of changes it would trigger, the "Singularity." He maintained that the process leading to the Singularity was irreversible, and that even the coordinated efforts of all the world's governments would be incapable of stop it.[4] Looking beyond this threshold, Vinge anticipated what he called the "Post-Human Era," characterized not by the end of humanity, but by the rise of superhumanity. In this context, the famous Asimov's Laws of Robotics would cease to be mere literary constructs of science fiction and become indispensable ethical frameworks for managing post-Singularity realities. To mitigate the risks of a negative outcome, Vinge proposed an alternative pathway: Intelligence Amplification. This would require not the development of pure artificial intelligence, but the nurturing of enhanced human intelligence through augmentation—a vision centered on the development of a mutually beneficial human-computer symbiosis, enabled by high-bandwidth networks.[5] Implicitly or explicitly, many scholars who theorize about the future apply a form of Darwinian logic to these transformations, even when not overtly acknowledged.

Kurzweil himself defines the Singularity as the culmination of exponential technological trends that are driving a transition poised to be profoundly transformative for humanity. This transformation spans several fundamental domains—including computing, human biology, and engineering—unified under the broader framework of Artificial Intelligence, or potentially even Superintelligence.[6] Another pundit who is dealing

---

[3] "Ulam, Stanislaw, "Tribute to John von Neumann," Bulletin of the American Mathematical Society. 64, (3), part 2:"1958, p. 5
[4] Vinge, Vernor, "The coming technological singularity: How to survive in the post-human era," NASA. Lewis Research Center, Vision 21, December 1, 1993, p. 14
[5] p. 20
[6] Ray Kurzweil, "The Singularity Is Nearer: When We Merge with AI," Penguin, 2024 p. 2

professionally with the AI and the idea of Singularity is Craig Wheeler, former president of the American Astronomical Society in his book he states that autonomous weapons "They could change warfare in ways that are very difficult to imagine."[7] And he adds that by referring to the drone warfare that "It is a small step from robot soccer to swarms of cooperating killer robots."[8] For him Singularity is inevitable process that can occur in our lifetime.

It is no coincidence that the concept of the Singularity originates in physics, particularly from the theory of black holes. In astrophysical terms, black holes are "singular" entities in space-time—regions where conventional physical laws collapse and cease to operate in predictable ways. By analogy, the technological Singularity is anticipated to trigger a phase of uncontrollable and accelerated technological advancement, resulting in a rupture with historical patterns of social and civilizational development. The normative frameworks and societal rules that have governed human life for millennia may no longer be applicable. As a result, the post-Singularity world is expected to be marked by radical unpredictability.[9] As writes Craig Wheeler that given all the revolutionary technologies in play and the factor of the acceleration of technological development, the singularity could actually be imminent, arriving during the lifetime of most people alive today or could lead to a singularity within months.[10]

The next element of my argument concerns the convergence of science fiction, rational projections of the future, and the tangible realities of the ongoing technological revolution. In the contemporary era, humans are increasingly capable of rationalizing the future with the same cognitive tools used to interpret the past and present. This rationalization, including of Euclidean space, reflects a broader process of the anthropogenic transformation of spacetime itself. The designation of the Anthropocene as a new geological epoch serves as direct evidence of humanity's transformative impact on the planet's biosphere. However, with the advent of cybernetics and non-biological intelligence systems, this transformation is poised to assume an unprecedented character. In this context, the role of science fiction and future-oriented scientific speculation is to outline potential trajectories for human history. As such, anthropic and cognitive factors begin to exert increasing influence over the biosphere—and, potentially, the noosphere. The latter concept, introduced by Russian and Ukrainian scientist Vladimir Vernadsky, refers to a planetary evolutionary stage in which human intellectual activity transforms the biosphere into a new form of life governed by reason and consciousness.

At the core of this argument lies the proposition that the future is becoming increasingly projectable. This may be described as a rationalization of the temporal continuum, wherein the "eclipse" of time becomes more controllable than ever before. Science fiction authors exemplify this phenomenon: they not only anticipate future developments but construct speculative frameworks for humanity's collective destiny—even when warning of potential catastrophes. In doing so, they create what might be termed a condition of "subjective objectivity," where imagined futures shape the real dynamics of technological and civilizational change.

---

[7] J. Craig Wheeler, "The Path to Singularity: How Technology Will Challenge the Future of Humanity," Rowman & Littlefield, 2024, p. 83
[8] Ibid., p. 84
[9] Susan Schneider, "Artificial You: AI and the Future of Your Mind," Princeton University Press, 2021, p.10
[10] Craig J. Wheeler, "The Path to Singularity: How Technology Will Challenge the Future of Humanity," Rowman & Littlefield, 2024, p. 5

The distance between imagination and realization—between prediction and reality—has never been as narrow as it is today, and available evidence suggests that this gap continues to shrink. While humanity may never attain complete control over the future, it is increasingly clear that we are living in an era where the future is becoming an extension of our cognitive imagination—one that materializes in physical and technological reality. This marks a new stage in human history. However, this development should not be interpreted as a guarantee of humanity's success or survival. On the contrary, the intensifying intersection of geopolitical rivalry and technological advancement introduces unprecedented risks for global stability and human security. The growing capacity to anticipate technological change does not equate to an ability to govern or direct the historical processes those changes unleash.

If we take into account that the term of the "Military-Technologic Revolution" which means that "…broader revolutions in warfare – periods of sharp, discontinuous change that render obsolete or subordinate existing military regimes or the most common means for conducting war. These changes may apply to militarily relevant technologies, concepts of operation, methods of organization, available resources, or a combination of several of these things. They are also often linked with broader political, social, economic and scientific transformations."[11] However, the advent of the Singularity transforms this episodic model into a condition of permanent revolution, wherein the military—an institutional descendant of the bureaucratic state apparatus of the eighteenth and nineteenth centuries—is increasingly unable to keep pace with the relentless accumulation of new technologies. This leads to an inflation of ideas without the capacity for timely and systematic integration. In the context of the Singularity, the state must develop new models of strategic assessment, particularly in the realm of net assessment. It is not coincidental that Ukraine has granted significant economic and institutional autonomy to its R&D sector and emerging tech start-ups—especially those working on drones, machine learning models, and related systems. The reduced involvement of the state apparatus in R&D has enhanced both the creative agility and production efficiency of these technologies in direct response to frontline demands. The Ukrainian model suggests a pragmatic refusal to invest in obsolete systems, emphasizing relevance and adaptability. For NATO, this presents a critical opportunity: by supporting Ukraine's military R&D ecosystem, the alliance can harness a new frontier in defense innovation, including the design of next-generation tanks, infantry fighting vehicles, and integrated battlefield systems. To date, Ukraine has demonstrated one of the most adaptive and responsive military-industrial complexes in the world. Notably, Russia has begun observing and adopting some of these emerging principles. In this sense, Ukraine may be understood as a potential cradle of Singularity Warfare—a site where the future of conflict is not only theorized but actively prototyped on the battlefield.

But it doesn't mean that the centralization is obsolete the Chinese approach towards the Singularity Warfare is unique in certain point because comparing with all traditional great powers it is only one that is closer than any to the "singularity" point. Already in 2016, the Chinese MoD stated that "On future battlefields, as AI and human-machine integration technologies advance, the tempo of combat will accelerate until it reaches a "singularity"—a tipping point at which the human brain can no longer cope with the rapidly changing battlefield dynamics and must relinquish most decision-making to highly intelligent machines. […] Ultimately, human soldiers may exit the kill chain, with intelligent machines becoming the primary combatants on future battlefields. Warfare will shift to a "human-on-the-loop" model in which humans mostly observe and occasionally intervene in machine-driven engagements."[12] The revolution is going to happen at every point of the war from soldier to the very concept of the Command and Control. The excellent American AI expert calls this process as "Intelligentized Command Decision-making" or simply "intelligentization."[13] Something similar has stated President of Russia and dictator by the describing the war with Ukraine and its technologic dynamics as following "Every month the conditions and methods of conducting armed struggle change: It is only necessary to fall behind for a few weeks - and the losses increase. Or the dynamics of advancement along the line of combat contact decreases. A few months is enough and that's it."

---

[11] Work, Robert O., and Shawn Brimley. 20YY: Preparing for War in the Robotic Age. Center for a New American Security, 2014, 7 // https://s3.us-east-1.amazonaws.com/documents/20YY_PreparingForWar.pdf

Furthermore, Paul Scharre from the Center for a New American Security contributes to the theory of the future warfare or if to use proposed by this paper concept the Singularity Warfare by stating that "…a revolution in warfare is under way driven in part by machine intelligence, then there is an imperative to invest heavily in AI, robotics, and automation. The consequences of falling behind could be disastrous for the United States. The industrial revolution led to machines that were stronger than humans, and the victors were those who best capitalized on that technology. Today's information revolution is leading to machines that are smarter and faster than humans. Tomorrow's victors will be those who best exploit AI."[14]

Thus, Paul Scharre introduces a crucial dimension to the technological transformation of civilization: the intensifying geopolitical competition among great powers, which serves as a primary driver accelerating humanity's entry into the era of the Singularity. Even if there were a collective desire to halt or slow the trajectory of history toward this transformative point, such a course is rendered effectively impossible. No state can afford to allow its adversaries to gain technological superiority, as doing so would risk undermining its own security—and potentially its very survival. In this context, what begins as a geopolitical necessity quickly evolves into a geopolitical imperative. This imperative exerts pressure on global centers of power to outpace their rivals in reaching and shaping the conditions of the Singularity.

Exactly, Scharre proposes the term the Battlefield Singularity, for him it is future battlefield with synergy of the swarming of the drones and robots under the command of AI. Scharre writes "Such an approach would be very challenging for humans to counteract, as it could overload the cognitive abilities of human defenders. If large-scale AI-driven swarming proved successful as an operational approach to organizing and employing military forces, other militaries could be forced to follow suit to survive. Such a development is likely decades away, if at all. There is a major leap between small tactical drone swarms, a near-term prospect,

---

[12] Chen Hanghui [陈航辉], "Artificial Intelligence: Disruptively Changing the Rules of the Game" [人工智能：颠覆性改变 "游戏规则], China Military Online, March 18, 2016, http://www.81.cn/jskj/2016-03/18/content_6966873_2.htm. Chen Hanghui is affiliated with the Nanjing Army Command College

[13] Kania, Elsa B. Battlefield Singularity: Artificial Intelligence, Military Revolution, and China's Future Military Power. Center for a New American Security (CNAS), November 2017, p. 5 // https://s3.amazonaws.com/files.cnas.org/documents/Battlefield-Singularity-November-2017.pdf

[14] Paul Scharre, "Army of None: Autonomous Weapons and the Future of War," W.W. Norton, 2018, p. 99

and the widespread use of AI-driven swarming across the entire battlefield. But AI could enable such a future."[15]

It can totally not only change the logic of war, but the Social Contract over the war because the era of Singularity is something that comparable with the Neolithic Revolution in the Human history or with Cambrian Explosion (the 'explosion' of the biodiversity) from the Earth history. The Neolithic Revolution, roughly 12,000 years ago, had enormous impact on our history it terraformed societal structure of the humanity from hunter-gatherer lifestyle to settled agricultural societies. As well as, it impacted on the nature of the warfare it became much more sophisticated and institutionalized during the Neolithic.[16] The American economist in this regards writes that the humanity already has witnessed at least singularities are Neolithic Revolution and Industrial Revolution and he adds that "So we have perhaps five eras during which the thing whose growth is at issue— the universe, brains, the hunting economy, the farming economy, and the industrial economy—doubled in size at fixed intervals."[17] These eras are result, according to him the process of the revolutionary innovations.

Or even something much greater I would refer to the to the prominent Polish futurist Stanislaw Lem who proposed the term of "necro-evolution" (Greek nekros, meaning "dead") that means the evolution of the non-biological substance or „a form of evolution of self-organizing mechanical systems."[18] Exactly about this writes Max Tegmark, who in his famous book "Life 3.0" proposes new dimension of the evolution on Earth. He defines life as the process that can retain its complexity and replicate or "self-replicating information-processing system whose information (software) determines both its behavior and the blueprints for its hardware."[19] Thus, we are witnessing evolution of Life from that stage 1.0 (simple biological), stge 2.0 (Cultural or Human) and Life 3.0 (Technological). This stage is only approaching according to Tegmark, but it is inevitable but he defines it as "master of its own destiny" would be able to write own software and hardware.[20]

As well as, Murray Shanahan a professor of Cognitive Robotics at Imperial College London goes even further in connecting the Technologic Singularity and the future warfare, "In wartime, military advantage might be had by allowing an artificial superintelligence to make near-instantaneous strategic and tactical decisions, both in the physical theater of operations and in cyberspace. The inherently competitive dynamics in these areas entail that if superintelligence could happen, it almost certainly would. For a corporation, the mere possibility that its competitors might obtain a decisive advantage by deploying machine superintelligence would be sufficient to ensure that it would try to get there first."[21] And he adds that "Despite this bleak assessment, the arguments in favor of the military use of AI are also worth attending to. Autonomous weapons are potentially more accurate and less error-prone than human combatants. They can be used more clinically, reducing so-called collateral

---

damage. Their decisions are never influenced by fear, revenge, or anger. […] But our present focus is not the rights and wrongs of military AI. The point is simply that the potential for military application is another driving force for the future development of sophisticated AI technology."[22] For him these transitions that are changing the very philosophy of warfare exactly the symptoms of the upcoming Singularity.

However, one special aspect of the Singularity — particularly as it manifests in warfare as the forefront of technological revolution - is the merging of human imagination with physical reality through technology. Plato spoke of intelligibility, of the world of ideas, which he called Eidos, existing apart from the material world. In modern warfare, this ancient duality dissolves: the cyber domain, virtuality, and algorithmic environments have become operational terrains where code becomes action, and imagination becomes executable logic.

Building on this foundation, one must include augmented cognition, autonomous decision-making systems, neuro-enhanced targeting, and synthetic environments as emergent layers where the Platonic realm of abstraction becomes tactically actionable. The battlefield is no longer only terrain-bound - it now includes cognitive architectures, data shadows, and semi-autonomous agents operating at machine speed.

It should be noted that this theory comes from the famous science historian Thomas Kuhn, who in his famous book "The Structure of the Scientific revolution" proposed one of the most profound visions on the understanding of the world history and philosophy of the science. The humanity according to Kuhn rises from one scientific paradigm to another and the process of the transition is occurring via the scientific revolutions. This process of happening by struggle of old and new paradigms. This struggle is not purely happening in the scientific cabinets or only in form technology but as well as the changes reverberate on the social structures and institutions. Kuhn in this regard's writes following "At that point, society becomes divided into competing camps or parties—one seeking to defend the old institutional constellation, and the other striving to institute a new one. This marks a moment of political and social crisis. Because the opposing sides differ fundamentally on the institutional matrix within which political change should be achieved and evaluated, and because they acknowledge no supra-institutional framework for the adjudication of these revolutionary differences, the parties are led into a revolutionary conflict. Eventually, they may resort to the techniques of mass persuasion, including force."[23] Kuhn dedicates several works to the quantum mechanics, but he describes the period that is known as a First Quantum Revolution. In the XXI century the humanity enters into the Second Quantum Revolution. Exactly the quantum mechanics that enabled development of technologies in the XX century such as lasers, the transistor, magnetic resonance imaging, and semiconductors, but there are the evidences that the surprises that we can expect from the quantum physics are paramount.[24]

Another important theory that underpins my argumentation was presented by the famous American futurologist Alvin Toffler. In the 90s Alvin and Heidi Toffler's 1993 book "War and Anti-War: Survival at the Dawn of the 21st Century." This idea contributed to the famous war doctrine known as the Network-Centric Warfare.[25] The core theme of the Toffler's work was that "the way we make war reflects the way we make wealth; and the way we make anti-war must reflect the way we make war"[26] Toffler's major idea is that the Human Civilization is undertakes grand cycles or waves of transformations that change model of their social, political life and the level of interaction with the nature. The humanity has witnessed two major waves that includes the Agrarian Revolution and Industrial Revolution. The latter started from the mid XVII century to the end of the XX century and was characterized by the high-level of the centralization, origination, and standardization. In the society it was presented itself as form of mass society, mass media, mass manufacturing, mass education etc.[27] But already in the 90s Toffler sees that the Humanity enters into new historical dynamics the Third Wave that will transform the society into the informational or intellectual form of the civilization. Also, it includes that the warfare that becomes the knowledge warfare "For decades to come, therefore, many of the best military minds will be assigned to the task of further defining the components of knowledge warfare, identifying their complex interrelationships, and building 'knowledge models' that yield strategic options. These will be the womb out of which full-blown knowledge strategies will be born."[28] He even suggests in 1993 that are going to emerge the knowledge warriors "- intellectuals in and out of uniform dedicated to the idea that knowledge can win, or prevent, wars."[29] However, it should be indicated that Toffler didn't idealize his notions he was certain that "We already know that older forms of warfare do not entirely disappear when newer ones arise. Just as Second Wave mass production has not disappeared with the coming of customized Third Wave products."[30] Finally, the real peak of the Third Wave is going to be when the robotization is going to occupy all spheres of the social action, for example, the warfare, Toffler in this regard adds that "Robots, like satellites and missiles and high-tech niche warfare, will, whether we are ready for them or not, take their place in the emerging war-form of Third Wave civilization."[31]

---

[22] Ibid., 154-155
[23] Kuhn, Thomas S., "The Structure of Scientific Revolutions," University of Chicago Press. 1996, p. 93
[24] Carson, J., "The Next Quantum Revolution," MIT Spectrum, Spring, 2024 // https://betterworld.mit.edu/spectrum/issues/2024-spring/the-next-quantum-revolution

      In this context, it is equally important to introduce a profound argument concerning the future of warfare—one that aligns closely with the conceptual framework of Singularity Warfare. German physicist Norbert Schwarzer, in his book Quantum Gravity Warfare, offers a theoretical and mathematical model suggesting that future conflict will be profoundly shaped by humanity's pursuit of a unified framework combining General Relativity and Quantum Theory. In the history of science, this effort is often referred to as the quest for a "Theory of Everything" or a Quantum Gravity Theory. Schwarzer's principal concern lies not merely in the theoretical elegance of such unification, but in its potential application within the military domain—a concern that echoes Thomas Kuhn's paradigm shift model. He argues that the realization of this scientific revolution would catalyze a series of overlapping technological

---

[25] Milan Vego, "Is the conduct of war a business,'" Army.mill., November 30, 2010 // https://www.army.mil/article/48803/is_the_conduct_of_war_a_business

[26] Alvin and Heidi Toffler, War and Anti-War: Survival at the Dawn of the 21st Century (Boston, MA: Little, Brown and Company, 1993), p. 2

[27] Interview with Alvin Toffler by Norman Swan, Radio National, March 5, 1998 // http://www.abc.net.au/rn/talks/lm/stories/s10440.htm

[28] Toffler, 1993, p. 199

[29] Ibid., p. 181

[30] Ibid., p. 107

[31] Ibid., p. 151

revolutions, unfolding simultaneously and fundamentally challenging classical understandings of science itself. In Schwarzer's view, information—not kinetic power—will become the defining resource in future warfare, with implications that stretch far beyond current paradigms.[32] There should be added that the String Theory is part of the great debate of the XXI century. In the Western academic debate already were voices that tried to connect the Fourth-Generation Warfare with the String Theory.[33]

Thus, we can conclude that in the recent decades has been developed non-mainstream scientific and philosophical vision or sort of the school that predicted the erection of the completely new form and principles than all previous scientific revolutions. To finalize my intellectual elucidations over future and Singularity particularly I need to turn to two fundamental scholars are Ilya Prigogine and David Deutsch. Ilya Prigogine in his book "The End of Certainty" (1996) presents the unique concept that challenging the Newtonian determinism. He writes that "Mankind is at a turning point, the beginning of a new rationality in which science is no longer identified with certitude and probability with ignorance. [...] In contrast, we believe that we are actually at the beginning of a new scientific era. We are observing the birth of a science that is no longer limited to idealized and simplified situations but reflects the complexity of the real world, a science that views us and our creativity as part of a fundamental trend present at all levels of nature." [34] The future science, according to him, will be determined by such characteristics like probabilities, irreversibility, fluctuations, bifurcations or chaos(*indeterminism*). If in the previous historic stages of the evolution of the science the great philosophers try to find the peace between new paradigm and reality, so in the future it would be hard achieving goal. Deutsch calls this 'appeasement' the "fabric of reality."

In Prigogine's physics the cognitive expressions like creativity, mind or culture becoming important components of the upcoming humanity's "turning point." For Prigogine the singularity is the "a point that contains the totality of the energy and matter in the universe."[35] The future isn't merely determinism it is constant bifurcation which almost impossible to predict. In this regard he wrote that "I am convinced that we are approaching a bifurcation point linked to progress in the development of information technologies and everything related to them, such as mass media, robotics, and artificial intelligence. This is the "networked society" with its dreams of a global village. […] My message to future generations, therefore, is that the die has not yet been cast, that the branch along which development will proceed after the bifurcation has not yet been chosen. We live in an era of fluctuations, when individual action still remains significant."[36]

Finally, David Deutsch, in his influential work The Fabric of Reality, offers a comprehensive framework for understanding our world through the lens of his Multiverse

---

[32] Norbert Schwarzer, "Quantum Gravity War: How Will the Nearby Unification of Physics Change the Future of Warfare," CRC Press, 2024, pp. 2-3

[33] "On New Wars," John Andreas Olsen (ed.), Oslo Files, 2007, p. 196 // https://fhs.brage.unit.no/fhs-xmlui/bitstream/handle/11250/99629/OF_4_2007.pdf?sequence=1&isAllowed=y

[34] Ilya Prigogine, "The End of Certainty," The Free Press, New York, 1997, p.7

[35] Ibid., p. 164

[36] Ilya Prigogine, „The Die Is Not Cast," Futures. Bulletin of the World Futures Studies Federation, Vol. 25, No.4. January 2000

theory. Grounded in quantum physics, Deutsch expands this model by integrating four foundational strands: quantum theory, Popperian epistemology (the theory of knowledge as conjecture and refutation), the Church–Turing–Deutsch principle (which posits that every physically possible process can be simulated by a universal computing device), and Darwinian evolutionary theory (as a general mechanism of adaptation). Through this synthesis, Deutsch proposes a new "fabric of reality" capable of explaining not only the current scientific landscape but also broader societal dynamics. Central to this paradigm is the quantum computer—a radically distinct entity from classical computing devices. Deutsch emphasizes that contemporary computers, regardless of their use of quantum effects, remain bound to the traditional model of the universal Turing machine. In contrast, the quantum computer represents an entirely new computational architecture, capable of simulating the multiverse and thus driving a deeper transformation in both science and society.

So, the future quantum computer "… is a machine that uses uniquely quantum-mechanical effects, especially interference, to perform wholly new types of computation that would be impossible, even in principle, on any Turing machine and hence on any classical computer. Quantum computation is therefore nothing less than a distinctively new way of harnessing nature."[37] By bringing these intellectuals as part of my theoretical argumentation I pursue one particular goal is to show that many outstanding intellectuals and scholars in the world in the recent decades have formulated their framework regarding upcoming scientific revolution that eventually is going to introduce completely new paradigm of reality. As result it will impact also on the warfare.

Let me present what I understand under proposed by me the term Singularity Warfare, it is universal, fluid and constantly changing evolution of forms and logic of warfare that integrates into the battlefield of all domains from space, sea, air, cyber etc. to cognitive. All currently developing cutting-edge-technologies (AI, language models, robotics, bio computing etc.) are eventually going to merge at the tip of the spear into new historical dynamics. We should understand that the current drones are only intermediary phenomenon from the perspective of technologic evolution and only miserable, but practically important piece of the broad concept of Singularity Warfare. Each of this domain will be facing the similar transformation of paradigm. It is something like if the rules of the chess game, the chessboard, chess pieces and even the players are changing at the same time. Everything is happening simultaneously and in one *horizon of the events* like in quantum physics when we have *superposition* (which a quantum system exists in multiple states at once) and under the condition of the *quantum entanglement* (two or more particles become linked so that the state of one instantly influences the state of the other). The human being simultaneously will be able to comprehend it in the cognitive capacity (similar to Kantian "Thing-in-itself" or "noumena"), but wouldn't be able absorb and processing the avalanche of the events comparing with evolving machines. It is going to be world of the cybernetically enhanced humans and AI-integrated robotics machines.

Thus, the war as social phenomenon permanently was undergoing paralleled changes as the process of the technologic revolution. This evolution simply impossible to decouple from the process of the scientific revolution and if we assume that we are at the stage of the Singularity Era and Quantum Revolution. It would definitely mean that the warfare undergoes the similar 'paradigm shift', but this time as was mentioned above that these changes are deeper and bigger in the time continuum. The humanity has used to the Newtonian type of the warfare it dominated over the human and its ecosystem almost entire period of existence of the human civilization. From point of militaries, we can call such type of warfare as the Clausewitzian notion that anticipates the war as a fundamental human endeavor.[38]

---

[37] David Deutsch, "The Fabric of Reality: The Science of Parallel Universes--and Its Implications," Penguin Publishing Group, 1997, p. 106

The war itself isn't source of transformation, but it is the source acceleration of the technologic and scientific revolution. The Singularity Warfare that incorporates the Quantum Revolution is accelerates as soon as the geopolitical global competition accelerates and it subsequently speed-ups the process of the societal transformation. It isn't coincidence that now quite perspective subject of research is the 'quantization' of the social sciences. According to Alexander Wendt, the quantum theory allows to settle several crucial long-standing anomalies in the social ontology: the existence of subjectivity (conscious aspect) and the unobservability of social structures.[39] He predicts that the evolution of the social structures and dynamics is moving from the classical physics to the "new physics," which is based mostly upon the foundations of the quantum mechanics. Wendt presents the future of the science as evolution from the decoherence to the quantum coherence, when the classical physics meets with the quantum principles.

In the context of Singularity Warfare, the decisive factor becomes the ability of the unit to maintain "coherence" - an integrated state of informational, cybernetic, cognitive, and combat systems (both human and robotic) in which every node in the network instantly responds to changes in any other. Similar to quantum coherence, this state enables the system to operate as a single organism, while the loss of coherence - "decoherence"—forces a reversion to fragmented, slower processes of classical warfare. Victory will be in the hands of those who can preserve coherence longer and disrupt it in the enemy faster. The enemy in Singularity Warfare will deliberately create "decoherence fields" – informational, cybernetic and psychological influences to destroy your coherence and return you to the "classical", slow response model. To have in the army the robotics and AI-systems at the big scale wouldn't be enough because if they wouldn't act in one coherency it would create potential danger and advantage for the enemy. For example, recently was published research on the topic of the quantum-cognitive tunnelling neural networks for military-civilian vehicle shows that we are witnessing the transition that marks a shift from traditional machine AI toward hybrid cognitive systems, where elements of human perception—uncertainty, context-dependence, cognitive biases—are embedded directly into neural architectures. It reflects the basic principles of the quantum physics like the superposition. One of the most perspective it has, according to the authors, in the drone warfare contexts, imbuing AI with the certain components of the human

---

[38] James Johnson, "Artificial Intelligence and the Future of Warfare: USA, China, and Strategic Stability," Manchester University Press, 2021, p. 3

[39] Alexander Wendt, "Quantum Mind and Social Science Unifying Physical and Social Ontology," Cambridge University Press, 2015, p. 7

reasoning and in the future, it can give significant operational advantage on the battlefield.[40] Figuratively speaking it is going to be stage when the human will be existing in the multiverse of our Newtonian reality and cybernetic universes, but the technologies will integrally part of these multiverses and intertwined between each other. I even would exclude the Clausewitzian warfare in the virtual universe between the great powers, non-state entities, AI inside these universes, but the results of these war would have impact on the physical reality. Again, I need to outline that I agree with point of Mark Sayers that we are still in the Gray Zone between the Singularity and Old World hence it is too early to think that the Singularity Warfare is going to conquer us immediately in this decade or so.

Within this theoretical framework, warfare is evolving into a phenomenon that increasingly resembles a video game—yet one situated in a state of quantum-like superposition, wherein commanders and soldiers operate simultaneously across three distinct but interconnected realities. The first can be described as the Newtonian reality—the physical and material domain governed by classical mechanics. It encompasses the tangible world of terrain, equipment, and bodily risk. The second is the virtual battlefield, a highly immersive and deeply layered environment whose dynamics will reverberate in the physical world. This digital space, characterized by near-infinite landscapes and procedurally generated architectures, will be designed not by humans but autonomously by AI. The third is an intermediary reality, in which human agents operate through embodied or disembodied systems—ranging from remote-controlled drones to mind-interface-operated platforms. In its early stages, this interaction may be mediated by joysticks and VR devices, but as neuro-interfaces advance, direct brain-machine integration may become the norm. These AI-enhanced, unmanned systems will increasingly assist or even supplant human decision-making, in our case for the commandeers and the soldiers, easing cognitive load while amplifying operational speed and scope. Collectively, these three layers suggest the emergence of a new ontological domain - one that may warrant the designation Synthosphere, in analogy to Vernadsky's Noosphere. This conceptual space would constitute a hybridized, synthetic reality where cognition, computation, and combat converge, fundamentally altering the nature of historical agency and strategic interaction that is natural for the humans.

**Conclusion**

Moreover, we don't know the effects of the Singularity on humanity and in our case on the theory and practice of the war. Therefore, it is so important in current geopolitical and techno-historical circumstances to every commander learn from the Russo-Ukrainian War wisdom of the synergy of utilization of old conventional forces with new forces and technologies which are forerunners of the Singularity Warfare.

This paper doesn't aim to produce the profound the principles and rules of the Singularity Warfare, but rather aims to open discussion on this issue. In the recent decades the intellectuals and scholars have formulated enormous number of the original theories and ideas regarding the future warfare, but unfortunately, they quite often only pointed to the skies but avoid theoretical framework that actually already was provided by the outstanding intellectuals like Thomas Kuhn, David Deutsch, Ilya Prigogine, Alvin Toffler, Isaac Azimov, Stanislaw Lem and etc. They exactly provided the theoretical and conceptual framework of the future, but my task was relatively hamble is to channel their knowledge and projections of the future into the warfare. I aware that my argumentation is weak particularly in the part of the warfare, but it wasn't direct goal of the paper because I thought I need to find the 'nodes' between the future warfare and theory of the future that is more historiosophic. I hope I managed to achieve this in this work.

---

[40] M Maksimovic, A Bohdanets, I Motsi-Omoijiade, G Governatori, IS Maksymov, "Quantum-Cognitive Tunnelling Neural Networks for Military-Civilian Vehicle Classification and Sentiment Analysis," arXiv preprint arXiv:2507.18645, 2025 // https://arxiv.org/abs/2507.18645

I would like to finish by referring to Frank Herbert, in his famous classical science-fiction "Dune" wrote extensively about the consequences of the AI for humanity. The catastrophic war with the "thinking machines" pushed Man for introduction of new commandment "Thou shalt not make a machine in the likeness of a human mind."[41] The rulers of the Emporium decided to halt any technological development that can create situation of preponderance of Man over Machine. According Lorenzo DiTommaso, Herbert in the Dune compartmentalized all aspects of technology into three main categories: the index of acceptable machines, the concentration of higher-order mathematics within the limited confines of the Spacing Guild, the virtual banishment of atomic energy (and probably research), and the continued use of millennia-old technological terminology. It was intentional decision to preserve the Man over the Machine, even was decided to return to the value of the human factor in combat ("personal combat and heroic elitism").[42]

---

[41] Stuart Russell, "Human Compatible: AI and the Problem of Control", Penguin UK, 2019 p. 237
[42] Lorenzo DiTommaso, „History and Historical Effect in Frank Herbert's Dune", Science Fiction Studies, 58, Volume 19, Part 3, November 1992 // https://www.depauw.edu/sfs/backissues/58/ditom58art.htm